\documentclass[psfig,preprint]{aastex}
\usepackage{amssymb,amsbsy}
\def\etal{{\it et al. \,}}

\def\be{\begin{equation}}
\def\ee{\end{equation}}
\def\bea{\begin{eqnarray}}
\def\eea{\end{eqnarray}}

\def\x{\boldsymbol{x}}
\def\v{\boldsymbol{v}}
\def\vnu{\boldsymbol{\nu}}
\def\n{\boldsymbol {n}}
\def\den{\sqrt{\Omega_{m0} a^{-3} + \lambda_0}}
\def\nden{\sqrt{\Omega_{m0} (\frac{\nu_e}{\nu})^{3} + \lambda_0}} 
\def\w{{ w}}
\def\la{\lower.5ex\hbox{$\; \buildrel < \over \sim \;$}}
\def\ga{\lower.5ex\hbox{$\; \buildrel > \over \sim \;$}}

\shorttitle{Using HI to probe large scale structures}
\shortauthors{Bharadwaj, Nath and Sethi}
\begin{document}
\title{Using HI to probe large scale structures at $z \sim 3$ .}
\author{Somnath Bharadwaj}
\affil{Department of Physics and Meteorology \& Center for Theoretical
Studies,\\
I.I.T. Kharagpur, 721 302, India}
\email{somnath@phy.iitkgp.ernet.in}
\author{Biman B. Nath}
\affil{Raman Research Institute, Bangalore 560 080, India}
\email{biman@rri.ernet.in}
\and
\author{Shiv K. Sethi} 
\affil{Harish-Chandra Research Institute, Chhatnag Road, Jhusi, Allahabad
211 019, India}  
\email{sethi@mri.ernet.in}

\begin{abstract}
 The redshifted $1420\, \rm MHz$ emission from the HI in unresolved damped
Lyman-$\alpha$ clouds at high z will appear as a background
radiation in low frequency radio observations. This holds the
possibility of a new tool for studying the universe at high-$z$, using
the mean brightness temperature to probe the HI content
and its fluctuations to probe the power spectrum. 
Existing estimates of the HI density at $z \sim 3$  imply a mean
brightness temperature of $1 \, {\rm mK}$ at $320 \, {\rm MHz}$.  
The cross-correlation between the temperature fluctuations across
different frequencies and sight lines is predicted to vary from $10^{-7}
\, \rm K^2$ to $10^{-8} \, \rm K^2$ over intervals corresponding to
spatial scales from $10 \, {\rm Mpc}$ to $40 \,  {\rm Mpc}$  
for some of the currently favoured cosmological models.  
Comparing this with the expected sensitivity of  the GMRT, we find that
this  can be detected with $\sim 10 \, \rm hrs$ of  integration,
provided  we can distinguish it from the galactic and extragalactic 
foregrounds which will  swamp this signal. We discuss a 
strategy based on the very distinct spectral properties of the
foregrounds as against the HI emission, possibly allowing the removal 
 of the foregrounds from the observed maps.  
\end{abstract} 
\section{Introduction}

The  problem of determining the distribution of matter on large scales
in the universe  and understanding the large scale structure (LSS) formation 
is of prime importance in modern cosmology. 
Observing the angular positions and redshifts of galaxies has
been the most straightforward method of mapping the LSS
in the present day universe (Peebles 1993, Peebles 1980), but the
interpretation of 
these observations  is complicated by the fact that the relation
between the distribution of  galaxies and the distribution of 
underlying matter is not fully understood (Bardeen \etal  1986).
Other direct methods
use galaxy clusters or super-clusters as tracers to  map the large
scale structures. A large variety of  techniques have been developed
and applied to quantify the distribution of galaxies, and   amongst these the 
correlation functions   (two-, three-point correlation functions, etc)
 and their Fourier counterparts (power spectrum, bi-spectrum, etc.)
have been most popular (Peebles 1980).  Much of the work comparing the 
observations with different theories has been based on
these.   

An alternative approach is  to use observations of fluctuations in the
diffuse
extragalactic background radiation at different  wavelengths to probe
the large scale structure. Here,  the  observations of
anisotropies in the  cosmic  microwave background radiation (CMBR)
have been most important. These observations probe the clustering of
matter at 
the last scattering surface (e.g. Bond 1996), and  combined with the
information from the galaxy surveys,  they have been successful in
narrowing down the allowed class of  theoretical models of LSS
formation  (e.g.,  Peacock 1999 and reference therein). The study of diffuse background
at other wavelengths has been used to infer the clustering properties
of matter at more recent epochs. 
Shectman (1974) observed  fluctuations in
the optical background; the results of this observation  are
consistent with predictions from
galaxy counts (Peebles 1980).  This method has also been
applied to far-infrared background (Kashlinsky {\it et al. \/} 1997)
and recently the first detection of fluctuations in  this background has
been reported  (Kashlinsky {\it et al. \/} 1999).
Gorjian, Wright and Chary (1999) have recently reported a tentative
detection of a near-infrared 
background. There are  similar predictions for 
fluctuations in the X-ray background (Barcons \etal 2000, Barcons \etal 1998).

In this paper we investigate the possibility of using the
extragalactic background radiation at low radio  frequencies (meter 
wave) to probe the LSS. 
This is largely motivated by the fact that  the Giant Meter-wave Radio
Telescope (GMRT; Swarup \etal 1991)  which is designed
to observe in  several frequency  bands in the  interval $150 \,{\rm
MHz}$ to $1420 \, {\rm MHz}$ has recently started functioning. 
This frequency range corresponds to the  $1420 {\rm MHz}$  emission
from HI in the redshift range $0 \le z \le 8.5$. Absorption studies
along sight lines to quasars indicate that most of the HI in the
redshift range $0 \le z \le 3$ is in damped Lyman-$\alpha$ (DLA)
clouds and the  density of HI  in DLAs has been 
determined reasonably  well  from absorption studies (Lanzetta,
Wolfe, \& Turnshek 1995). Although the HI emission from individual DLAs
at high $z$ is too faint to detect using presently available
telescopes, the redshifted HI emission from unresolved DLAs will
contribute  to the  background radiation at low frequencies. In this
paper we investigate how the brightness temperature of this radiation
is related to the density and peculiar velocity of the HI.  We
consider  the possibility of detecting correlations in the
fluctuations in this component of the background radiation and using
this to probe LSS at high $z$. 

The possibility of observing the redshifted 21 cm emission from HI at high
$z$ has been discussed earlier by many authors in a large variety of
contexts. One of the first papers (Sunyaev \& Zel'dovich, 1975)
considers the possibility of meter-wave observations to detect
protogalaxies and protoclusters at $z \le 10$. There have 
been several attempts to detect the HI in proto-clusters and 
proto-super clusters (see Subrahmanyan \& Anantharamaiah 1990 and 
reference therein).

 More recently 
Subramanian and Padmanabhan (1993) have calculated the abundance of
protoclusters as a function of the redshifted HI  flux density for
various redshifts 
for both the CDM and HDM models. In a later paper Kumar, Padmanabhan and
Subramanian (1995) have calculated the line profile of the HI emission
from a spherically symmetric protocluster.  Bagla, Nath and Padmanabhan
(1997) and Bagla (1998) have used a combination of  N-body simulations
and a  model for the behaviour of the baryons to
calculate the abundance and the expected flux from the HI in
structures like protoclusters at high redshifts. A major uncertainty
in all of these works is in the assumptions about the HI content of
the universe at high redshifts.
 The main focus of all of these works has been on 
individual peaks of the density fluctuations (protoclusters)  
which will manifest themselves as {
detectable features  in low-frequency  
radio maps. Subramanian and Padmanabhan (1993) have also studied the
possibility of detecting the excess variance in radio images due to
the density fluctuations in the HI at high $z$.

Katz, Weinberg and Hernquist (1998)  have used  smoothed
particle hydrodynamic simulations to study the distribution of HI gas
at high $z$ and they consider  the possibility of detecting HI
emission from galaxies at $z>2$.

Tozzi \etal (1999) and Shaver \etal (1999) have studied the
possibility of detecting the HI in the IGM at $z>5$. 
The state of the HI at these redshifts is  unknown and these works are
based on different scenarios for the reionization of the universe. 

The work presented here differs significantly from the previous papers
in that:
\begin{itemize}
\item[(1)] It is restricted to $z<3.5$ where the HI content of the
universe is well determined from DLA absorption studies. We use the
results of these observations as inputs to our calculations. 
\item[(2)] Instead of looking at the possibility of detecting
individual features (as has been the focus of a large number of
previous papers) we have studied the statistical properties of the
fluctuations in the brightness temperature in low frequency radio
maps. The statistical quantity we have studied  is the cross-correlation
between the temperature fluctuations along different lines of sight in
radio map made at different frequencies. Individual features
corresponding to protoclusters are rare events and protoclusters
with flux in the range $1.5-3 \, {\rm m \, Jy}$ are predicted to occur
with abundances in the  range $10^{-8} - 10^{-7} \, {\rm Mpc}^{-3}$ in
the CDM model  (Subramanian and Padmanabhan, 1993).
 Even small density
fluctuations will contribute to the cross-correlation signal and our 
proposal has the advantage of simultaneously using the signal in 
all the pixels in all the frequency channels across the bandwidth of
observation. 
\end{itemize}

The structure of the paper is as follows: in \S 2 we discuss
the basic formalism of calculating  the brightness temperature and
fluctuations in the redshifted 
21 cm radiation, in \S 3 we discuss the numerical results for two
currently favoured cosmological models. In \S 4, the observational
difficulties in the presence galactic and extragalactic foregrounds
are  presented, and we discuss a possible strategy for overcoming
these. \S 5 gives a summary of our main results.

\section{Formalism.}
We treat the HI in damped Lyman-$\alpha$ (DLA) systems as a
continuous distribution  with $n_{\rm HI}(\x,t)$ denoting the comoving
number density of HI atoms in the excited  state of the
hyperfine transition. Such a treatment is justified in a situation
where the resolution of the observations is not sufficient to detect
individual DLAs. In addition, the fact that the HI actually does not
have a continuous distribution but is  distributed in discrete objects
can, if required.  be taken into account when calculating statistical
measures of the fluctuation in the HI distribution.

The  HI emission which is at a frequency $\nu_e=1420 {\rm MHz} $  in
the rest frame of the gas it is emitted from   is redshifted to a
frequency $\nu$ for an 
observer located at the origin of the coordinate system. Taking into
account the effects of both the expansion of the universe as well as
the peculiar velocity  $\v(\x,t)$ of the HI  at the time of emission
and  $\v(0,t_0)$ of the observer at the present time,
the comoving coordinate $\x$  of the HI and $\nu$ are related by 
\be
\x= \n \,c \int_{\frac{\nu}{\nu_e  (1-W)}}^{1} \frac{d \, a}{a^2 H(a)}
\label{eq:a1}
\ee
in a spatially flat universe. Here $\n=\x/x$ is a unit vector along
the line of sight to the HI, 
$W(\x)=\n \cdot [\v(\x,t)-\v(0,t_0)]/c$ accounts for the effects of
the peculiar velocities and 
\be
H(a)=\frac{\dot{a}(t)}{a(t)}=H_0 \den
\ee
is the Hubble parameter at the epoch when the scale factor has value $a$. 

When discussing observations of the redshifted HI emission,  it is
convenient to use ${\vnu} = \n \nu $ to simultaneously denote the
frequency and the 
direction of the observation.  The vector $\vnu$ fixes both the
comoving position $\x$ of the HI from where the radiation originates
and the time $t$ at which the radiation originates, and we shall use
$\vnu$ and $(\x,t)$  interchangeably. Here we  
calculate how $T(\vnu)$ the brightness  temperature of the radiation  is
related to the density $n_{\rm HI}(\x,t)$ and the the peculiar
velocity $\v(\x,t)$ of the HI.  

The energy flux in the frequency interval $d^3 \nu$ can be
calculated if we know the number of excited HI atoms in the comoving
volume $d^3 x$  from where the radiation originates and it is given by
\be
{\rm Flux} =  \frac{ h_P \,\nu_e \, A_{21}\,  n_{\rm HI}(\x,t)\,  d^3 x}
{4 \,\pi \,  r^2_L(\x)} 
\label{eq:a2}
\ee
Here  $r_L$ is the luminosity distance which is given by 
\be
r_L(\x)=x \, [1-W] \, \frac{\nu_e }{\nu }
\ee

The flux is also related to the specific intensity as follows 
\be
{\rm Flux}=\frac{I(\vnu)}{\nu^2} d^3 \nu 
\ee
which allows us to calculate the specific intensity to be 
\be
\frac{I(\vnu)}{\nu^2}= \frac{ h_P \,\nu_e \, A_{21}\,  n_{\rm HI}(\vnu)}
{4 \,\pi\nu_e^2} \left\{ \frac{1}{[1-W(\vnu)]} \frac{\nu} {x} \right\}^2
\left| \frac{\partial \x }{\partial \vnu} \right| \,.
\label{eq:a3}
\ee
where $| \frac{\partial \x }{\partial \vnu} |$ is
the  Jacobian of the transformation from $\vnu$ to $\x$ given in
equation (\ref{eq:a1}). 

It should be noted here that the effect of the peculiar velocity
$(i.e. W)$ can be neglected if we restrict the analysis to scales
which are much smaller than the horizon. This is not true for the
spatial  derivatives of the peculiar velocities which appears in the
Jacobian  and we retain such terms  in our analysis. 
Calculating the Jacobian gives us
\be
\left | \frac{\partial \x }{\partial \vnu} \right | = 
\left( \frac{x}{\nu} \right)^2 \frac{c}{H(\frac{\nu}{\nu_e})}
\frac{\nu_e}{\nu^2}  \left[ 1 -
\frac{\nu_e}{\nu} \frac{(\n \cdot \boldsymbol{\nabla}) (\n \cdot \v)}
{H(\frac{\nu}{\nu_e})} 
 \right] \,.
\label{eq:a4}
\ee

Equations (\ref{eq:a3}) and (\ref{eq:a4})  allows us to calculate the
brightness temperature of the radiation 
\be
T(\vnu)= \left( \frac{c^2}{2 k_B} \right) \frac{I(\vnu)}{\nu^2}
\ee
which gives us 
\be
T(\vnu)=\frac{T_{21} N_{21}(\vnu)}{8 \pi}
\frac{A_{21}}{H(\frac{\nu}{\nu_e})}   \left(\frac{\nu_e}{\nu}\right)^2
\left[ 1 - \frac{\nu_e}{\nu} \frac{(\n \cdot \boldsymbol{\nabla}) (\n \cdot \v)}
{H(\frac{\nu}{\nu_e})} \right] \,.
\label{eq:a5}
\ee
where   $T_{21}=h_p \nu_e/k_B$ and $N_{21}(\vnu)=(c/\nu_e)^3 \, n_{HI}(\vnu) $ 
is the number of HI atoms in 
the excited state in a comoving volume  $(21 {\rm cm})^3$.

The number density $n_{\rm HI}(\vnu)$ can be written as a sum of two
parts, namely the mean $\bar{n}_{HI}(\nu)$ and the fluctuation 
$\Delta n_{HI}(\vnu)$. We assume that the fluctuation in the number
density of HI can be related to the perturbations in the underlying
dark matter distribution  $\delta(\vnu)$  through a time dependent
linear bias parameter $b(\nu)$ which gives us    
\be
N_{21}(\vnu)= \bar{N}_{21}(\nu)[1 + b(\nu) \delta(\vnu)] \,.
\ee
We use this to calculate the isotropic part of the  temperature 
\be
\bar{T}(\nu)=\frac{ 2.38\, {\rm K} h^{-1}\, \bar{N}_{21}(\nu)}{\nden}
\left(\frac{\nu_e}{\nu}\right)^2
\label{eq:a6}
\ee
and the fluctuation 
\be
\Delta T(\vnu)=\bar{T}(\nu) \left[ b(\nu) \delta(\vnu) -
\frac{\nu_e}{\nu} 
\frac{(\n \cdot \boldsymbol{\nabla}) (\n \cdot \v)}
{H(\frac{\nu}{\nu_e})} \right]
\label{eq:a7}
\ee

Now, in the linear theory of density perturbations, both the
perturbation and the peculiar velocity can be expressed in terms of a
potential i.e.
\be
\delta(\x,t)=D(t)\nabla^2 \psi(\x)
\ee
and 
\be
\v(\x,t)=-f(\Omega_{m}) H(t) a(t) D(t) \bf{\nabla} \psi(\x) \,.
\label{eq:a8}
\ee 
where $D(t)$ is the growing mode of linear density perturbations
(Peebles 1980). 
In a spatially flat universe the function $f$ can be well
approximated by the form  (Lahav et. al. 1991)
\be
f(\Omega_m)=\Omega_m^{0.6}+\frac{1}{70}[1-\frac{1}{2}\Omega_m(1+\Omega_m)]
\ee
where the time dependence of $\Omega_m$ can be expressed as  
\be
\Omega_m(\nu)=\left(\frac{H_0}{H(\frac{\nu}{\nu_e})}
\right)^2 \left(\frac{\nu_e}{\nu}\right)^3 \Omega_{m0} \,.
\ee
Using this and defining a time dependent linear redshift space
distortion parameter $\beta(\nu)=f(\Omega_m)/b(\nu)$ we can express
the fluctuation in the temperature as 

\be
\Delta T(\vnu)=\bar{T}_A(\nu) b(\nu) D(\nu)\left[   \nabla^2  +
\beta(\nu) (\n \cdot \bf{\nabla})^2 \right] \psi(\x) 
\label{eq:a10}
\ee

We next calculate the cross-correlation between  the temperature
fluctuations along two different lines of sight $\n_1$ and $\n_2$ at
two different frequencies $\nu_1$ and $\nu_2$. 
The  quantity we consider is the correlation
\be
\w(\vnu_1,\vnu_2) = \langle \Delta T(\vnu_1) \Delta T(\vnu_2) \rangle
\ee
which is a function of the two frequencies $\nu_1$, $\nu_2$ and 
$\theta$ the angle between the two lines of sight. 
Using equation (\ref{eq:a10}) and defining 
$\phi(x_{12}) =\langle \psi(\x_1)\psi(\x_2) \rangle$  to be the two point
correlation of the potential $\psi(\x)$, we  obtain
\be
\w(\vnu_1,\vnu_2)=\bar{T}_1 \bar{T}_2 D_1 D_2
b_1 b_2 \left[   \nabla^2  + \beta_1 (\n_1 \cdot
\boldsymbol{\nabla})^2 \right] \left[   \nabla^2  + \beta_2 (\n_2 \cdot
\boldsymbol{\nabla})^2 \right] \phi(x_{12})
\label{eq:a11}
\ee
where $x_{12}=\mid \x_1-\x_2\mid$ and we have used the notation
$\bar{T}_1=\bar{T}(\nu_1)$, etc.  

The function $\phi(x)$ is related to the correlation of the
perturbations in the underlying dark matter distribution, and 
$\xi(\x_1,t_1,\x_2,t_2)= \langle \delta(\x_1,t_1) \delta(\x_2,t_2)
\rangle$ the two point correlation  of the perturbation in the dark
matter density at the point $\x_1$ at the  epoch $t_1$ and $\x_2$ at
the epoch $t_2$ can be written in terms of the  potential $\phi$ as 
\be
\xi(\x_1,t_1,\x_2,t_2)=
D(t_1) D(t_2) \tilde{\xi}(x_{12}) 
=D(t_1) D(t_2) \nabla^4 \phi(x_{12}) \,.  \label{eq:a12}
\ee
We have introduced the function
$\tilde{\xi}(x_{12})$ so that we can write $\xi(\x_1,t_1,\x_2,t_2)$
 as a product of two parts, one which has the temporal variation and
another the spatial variation.  Equation (\ref{eq:a12}) can be
inverted to express the different derivatives of $\phi(x_{12})$ which
appear in equation (\ref{eq:a11}) in terms of  moments of
$\tilde{\xi}(x_{12})$ which are defined as 
\be
\tilde{\xi}_n(x_{12})=\frac{n+1}{x^{n+1}_{12}} \int^{x_{12}}_0
\tilde{\xi}(y)
y^n d \, y \,.
\ee
The form of the angular correlation is further simplified if we
restrict 
$\theta$ to be very small. Under this assumption  $\n_1
\simeq \n_2$,  and we use $\n$ to denote the common line of sight. We
also use $\mu=\n \cdot (\x_1-\x_2)/x_{12}$ for the cosine of the angle between
the line of sight $\n$ and the vector $\x_1-\x_2$ joining the two
points between which we are measuring the correlation. The relation
between the different distances and angles is shown in figure
\ref{fig:1}, and we  have $x_{12}=\sqrt{x_1^2 + x_2^2- 2 x_1 x_2
\cos(\theta)}$ and $\mu=(x_1 - x_2 \cos(\theta))/x_{12}$.       
Using these we finally obtain the following  expression for the 
two point correlation of the temperature 
\bea 
\w(\nu_1,\nu_2,\theta)&=& \bar{T}_1 \bar{T}_2 D_1 D_2
b_2 b_2 \left\{ \left[(1+ \beta_1 \mu^2)(1+ \beta_2 \mu^2)
 \right] \tilde{\xi}(x_{12}) \right. \nonumber \\
&+& \left. \left[(\frac{1}{3} -\mu^2) (\beta_1 + \beta_2) + (\frac{1}{2}-3
\mu^2  + \frac{5}{2} \mu^4) \beta_1 \beta_2 \right]
\tilde{\xi_2}(x_{12}) \nonumber \right. \\
&-& \left. \left[ \frac{3}{10} + 3 \mu^2 -\frac{7}{2} \mu^4 \right] 
\tilde{\xi_4}(x_{12}) \right\}
\eea

\section{Predictions for different models.}
The  density of HI ($\Omega_{\rm HI}$) in DLAs has been
determined reasonably  well  for  $0 \le z \le 3$  from absorption
studies (Lanzetta, Wolfe, \& Turnshek 1995)  and they find that the
observed evolution of $\Omega_{\rm HI}$ is well approximated by 
$ \Omega_{\rm HI}(z)=\Omega_{\rm HI0} \exp(\alpha \, z) $ with 
$\Omega_{\rm HI0}=0.18 \pm 0.04 \times 10^{-3} h^{-1}$ and
$\alpha=0.60     \pm 0.15$ for $q_0=0$, and
$\Omega_{\rm HI0}=0.19 \pm 0.04 \times 10^{-3} h^{-1}$ and
$\alpha=0.83     \pm 0.15$ for $q_0=0.5$, We have used this to
calculate the isotropic part of the  background brightness temperature
due to  the
redshifted HI  emission (equation \ref{eq:a10}). 
We have considered two spatially flat FRW cosmological models with
parameters  (I) $\Omega_{m0}=1.0$ and $\lambda_0=0$, and (II) 
$\Omega_{m0}=0.3$ and $\lambda_0=0.7$, with $h=0.5$ in both cases. 
It should be pointed out that the fits given Lanzetta {\it et
al. \/} (1995) are not valid for the model with a cosmological
constant, and  we use the $q_0 = 0$ fit in this case.  

Figure \ref{fig:2}  shows the brightness temperature  as a function
of frequency for both the cases. 

We find that the temperature increases
rapidly as we go to lower frequencies (higher $z$) and it is around $1 \,
{\rm mK}$ at $\nu \sim 330 {\rm MHz}$ which corresponds to $z \sim
3$. The increase in the temperature is a direct consequence of the
increase of HI density with increasing redshift. The HI density is not
very well determined at $z > 3$ and  there is evidence
that $\Omega_{\rm HI}$  falls off at higher redshifts
(Storrie--Lombardi {\it et al. \/} 1996).

We next consider $\w(\nu_1,\nu_2,\theta)$ the cross-correlation in the
temperature fluctuation  at different frequencies which we have
calculated  for the two cases discussed above. We have calculated the
dark-matter two point correlation 
function (equation \ref{eq:a12}) using the analytic fitting form
for the CDM power spectrum given by Efstathiou, Bond and White
(1992). For case (I) we use a value of the shape parameter
$\Gamma=0.25$ and for (II) we use $\Gamma=0.14$. The
power spectra are  normalized using the results of  Bunn and  White (1996)
based on the 4-year COBE data. 
It should be noted that he redshift evolution of the matter
correlation function and $f(\Omega_{m})$ are quite  different in the
two cases that we have considered. 
The predictions for $\w(\nu_1,\nu_2,\theta)$ are shown in figures
\ref{fig:3} and \ref{fig:4}.  We have assumed that the HI faithfully
traces the matter distribution and set $b(\nu)=1$ throughout. 

	In our analysis we have kept $\nu_1$ fixed at $320 {\rm MHz}$
and let $\nu_2$  vary over a band of $16 {\rm MHz}$ centered
around $\nu_1$, while $\theta$ takes values upto $2^{\circ}$. 
In order to  measure $\w(\nu_1,\nu_2,\theta)$ over this range we would
need radio images of the background temperature fluctuation in a
$2^{\circ} \times 2^{\circ}$  field at  different frequencies in a
$16 {\rm MHz}$  band centered around $320 {\rm MHz}$ and 
$\w(\nu_1,\nu_2,\theta)$ could be estimated using
\be
\w(\nu_1,\nu_2,\theta)=\langle \Delta T_{\nu_1}(\n_1) \Delta
T_{\nu_2}(\n_2) \rangle
\ee
where $\Delta T_{\nu_1}(\n_1)$ refers to  the temperature fluctuation
along  the direction $\n_1$ in the image made at frequency $\nu_1$,
and the angular brackets denote average over all pairs of directions
$\n_1$ and $\n_2$ which are separated by an angle $\theta$. 
The central frequency, bandwidth and angular range have been chosen
keeping in mind the Giant Meter wave Radio Telescope (GMRT, Swarup
\etal 1991) which has recently become operational.  The frequency
intervals and angular separation can be converted to a corresponding
comoving length-scale and in case  (I) 
$1 {\rm MHz}=8.9 {\rm Mpc}$ and $1^{'}=1.8 {\rm Mpc}$, while in case
(II)  $1 {\rm MHz}=16.0 {\rm Mpc}$ and $1^{'}=2.7 {\rm Mpc}$.  The
linear theory of density perturbations which we have used here can be
applied at  scales $\ge 10 {\rm Mpc}$, which covers most of the region
shown in the figures.  

We find that $\w(\nu_1,\nu_2,\theta)$ is between $10^{-7} {\rm K}^2$  
to $10^{-8} {\rm K}^2$  when the comoving-distance corresponding to
the separation in frequencies and direction is less between $10 {\rm
Mpc}$ and  $40 {\rm Mpc}$ beyond which it falls off. The
cross-correlation between the temperature fluctuations 
is expected to be larger at  small scales where the linear theory of
density perturbations will not be valid and we have not considered
these scales 
here.  The temperature fluctuations are anti-correlated when the
separation in  frequencies exceeds the angular separation. This occurs
because of the effect of the peculiar velocity which produces a
``distortion'' very similar to the effect it has on the two-point
correlation function in redshift surveys.  

\section{Observational prospects}
In this section  we discuss the prospects of actually observing the
background radiation from the HI in damped Lyman-$\alpha$ clouds. Our
discussion is restricted to observations at around $320 \, {\rm
MHz}$ largely because we have reliable estimates of the HI content
at $z \le3$ and there are indications that $z \simeq  3$  might be the
redshift where the HI content of the universe is maximum
(Storrie--Lombardi {\it et al. \/} 1996).
Another reason for our choice of this 
frequency range  is that the GMRT (Swarup \etal 1991) which is already 
functioning at these frequencies is expected to have an angular
resolution of $10^{''}$ and reach noise levels of $100
{\rm \mu K}$ in around 10 hrs of integration.
The angular resolution and sensitivity of the GMRT will be
sufficient for detecting both the isotropic component as well as the
correlations in the component of the background radiation arising from
the HI at $z \simeq 3$ (Figures~\ref{fig:2},~\ref{fig:3}
and~\ref{fig:4}) provided we can distinguish this component  
from the contribution due to other sources. 

\subsection{Galactic and extra-galactic foregrounds}

Any observations at low frequencies will have a very large contribution
from the  synchrotron radiation from our own galaxy. Observations
at $408\, \rm MHz$ (Haslam {\it et al. \/} 1982) with resolution 
$1^\circ \hbox{x}1^\circ$ indicate a minimum temperature of  $\simeq
10 \, \rm K$ at $408  \,{\rm MH}z$. Using $T \propto \nu^{-2.7}$ as
indicated by the spectral 
index of galactic synchrotron radiation, the temperature at $\nu
\simeq 320 \, \rm MHz$ is $\simeq 20 \, \rm K$. Comparing this
number with the expected background from redshifted HI radiation
(Figure~\ref{fig:2}) the galactic foreground is seen to be several
orders above the expected signal. Another quantity of
interest to us is the fluctuations in the galactic radiation at angles $\la
1^\circ$. However as the resolution of
Haslam maps is $\simeq 1^\circ$ it cannot be used to make any
predictions about the fluctuations at the angular scales of interest.

Another source of contamination is  the continuum radiation from
unresolved  extragalactic sources  (the resolved ones will be removed
from the image before  analysis) and we use results from the recent
FIRST survey (for details see White {\it et al.\/} 1997 and references
therein)  to estimate this. Tegmark \& Estathiou (1995) provide an
analytic fit for  the number of sources  per unit flux
$\phi[{\rm mJy}]$ per steradian in this survey, and at flux levels $\phi
\ll 100 {\rm  mJy}$ this can be approximated by  
\begin{equation}
{dn \over d\phi} = {5.24 \times 10^5 \over {\rm mJy \, sr}} \left
  ({\phi \over 0.75 {\rm mJy}} \right )^{-1.65} \,.
\label{eq:foreg}
\end{equation}

We use this to estimate the total contribution from sources
fainter  than $100 \, \mu {\rm Jy}$, and converting this to brightness
temperature  we find that the continuum emission from unresolved
radio sources  is expected to  produce a background radiation with
brightness temperature $\sim .1 \rm K$ at $1.5 \,  {\rm Ghz}$.  
For the purposes of estimating  an order-of-magnitude  we
assume this  value to be representative of what we expect at  $\sim
320 \, \rm MHz$.

The two-point correlation function of the sources detected by FIRST
has been estimated (Cress {\it et al. \/} 1996): $w(\theta) \simeq
0.18 \theta^{-1.1}$, with $\theta$ in $\rm arcminutes$. Little is
known about the correlations expected in the fluctuations in the
contribution to the background radiation from unresolved radio
sources.  

It is clear from the foregoing discussion that both galactic and
extragalactic continuum signals are likely to be so high that they
would totally swamp the signal which we want to detect  and  therefore
it is not possible to  directly  detect the  radiation from  HI unless
we find some method for removing the foregrounds.

 At present the most promising strategy is to use the proposed
observations themselves to determine the foreground and remove this
from the data, and we next discuss a possible method for doing this.

\subsection{Removing  foregrounds}

We consider the GMRT band centered at  $325 \, \rm MHz$ for our
discussion in this section. The total bandwidth of $16 \,  \rm MHz$ is
divided in 128 
channels each with $\Delta \nu = 1.25 \, \rm kHz $, and   the signal in any
frequency channel is expected to be dominated  by the foregrounds. 

Here we briefly review the standard method of continuum subtraction
which is generally used in spectral line detections and then discuss
how this method can be used for the  analysis proposed here. The
reader is 
referred to Subrahmanyam and Anantharamaiah (1990) and references
therein for examples of how this method is applied in searches for HI 
emission from protoclusters. 

We represent the  observed signal $y_i(\n)$ in the $i$ th frequency
channel in the direction $\n$ as:
\begin{equation}
 y_i(\n) = x_i(\n) + f_i(\n) + N_i 
\end{equation}
where $x_i(\n)$ and $f_i(\n)$ are  the contribution to the output
signal from the HI and  the foregrounds respectively,  and $N_i$ is
the  receiver noise. 

We should at this stage remind ourselves that  foregrounds
are sources which are emitting continuum radiation---the primary such
source being the synchrotron radiation from our own galaxy. We also
have the contribution from unresolved extragalactic radio sources
which, as we have seen above, is a smaller contribution. 
The sum total of these along any line
of sight is expected to be a smooth function of the frequency. On the
contrary the contribution from the HI will come from individual damped
Lyman $\alpha$ clouds along the line of sight with a velocity width
$\simeq 200 \, \rm km \, sec^{-1}$ (Prochaska \& Wolfe 1998) and this will
correspond to lines of width: 
\begin{equation}
\Delta \nu \sim \frac{\Delta V}{c} \nu \simeq 0.2 {\rm MHz}.
\end{equation}

Although the total number of damped Lyman-$\alpha$ clouds in a
$1^{\circ} \times 1^{\circ} \times \, 16 \, {\rm MHz}$ field is expected to be
quite large $(\simeq  3 \times 10^4)$ we expect to have only one damped
Lyman-$\alpha$ cloud along a single $20^{''} \times 20^{''}
\times  16 \,{\rm MHz}$ synthesized GMRT beam. 
The probability of two damped Lyman-$\alpha$ clouds occurring right
next to each other in the same synthesized beam  and causing confusion
is quite  small,  and hence the signal we are looking for will come in
the form of  lines of width $\sim 0.2 {\rm MHz}$ which we will not be
able to 
detect as the individual lines will be swamped by both the
foreground as well as the noise. What we  can do is to fit the smooth
component of  the signal $y_i(\n)$  by a  function $F_i(\n)$ and
subtract this from the output 
signal to remove the effect of the foregrounds  
\begin{equation}
S_i(\n)=y_i(\n)-F_i(\n)=x_i(\n)+N_i(\n)+E_i(\n)
\end{equation}
where we call $S_i(\n)$ the reduced signal and $E_i(\n)$ is the
possible error in the fitting procedure. If our foreground subtraction
works correctly, the reduced signal should have contributions from
only the HI signal and the noise.

We first determine the mean reduced signal 
\begin{equation}
\bar{S}_i=\left \langle S_i(\n) \right \rangle   = \left \langle
  x_i(\n) \right \rangle +\left \langle
N_i(\n) \right \rangle + \left \langle E_i(\n) \right \rangle 
\end{equation}
where the angular brackets  denote average over all lines of
sight. If the errors in the foreground  subtraction can be made
smaller than the signal i.e. $\langle E_i(\n) \rangle  \, < \, 
\langle x_i(\n)\rangle$
then this will give an estimate of the background
brightness temperature due to the HI emission
i.e. $\bar{T}(\nu)=\bar{S}_i$ where the index $i$ refers to the
channel with central frequency $\nu$.

We next consider the quantity
\begin{equation}
 \Delta S_i(\n)=S_i(\n)-\bar{S}_i
\end{equation}
which  gives the fluctuation in the brightness  temperature. 
We can use this to estimate the cross-correlation in the fluctuations
at different frequencies 
\begin{equation}
\w(\nu_1,\nu_2,\theta)=\langle \Delta S_i(\n) \Delta S_j({\bf m}) \rangle
\end{equation}
where $i$ and $j$ are  the channel with frequency $\nu_1$ and $\nu_2$
respectively, and $\n$ and ${\bf m}$ are two lines of sight separated
by an angle $\theta$ and the angular brackets denote average over all
such pairs of lines of sight.

\section{Summary and Conclusions}
We have investigated the contribution from the HI in unresolved damped
Lyman-$\alpha$ clouds at high redshifts to the background radiation
at low frequency radio waves (meter waves). The isotropic part of this
radiation depends on the density of HI and the background cosmological
model, while the fluctuations in this component of the background
radiation have an added dependence on the fluctuations in the distribution
of the damped Lyman-$\alpha$ clouds and their peculiar velocities.   

We have used estimates of the HI density available from absorption
studies to calculate the brightness temperature of this radiation. We
find that this has a value $\sim 1 \, {\rm mK}$ at $320 \, {\rm MHz} $
which corresponds to $z \sim 3$. The  distribution of damped
Lyman-$\alpha$  clouds is assumed to trace the underlying dark matter
distribution  which also determines the peculiar velocities. Using
this and the  linear  theory of density perturbations, we have
calculated the relation between the fluctuation in this 
component of the background radiation and the density perturbations at
high $z$. Observations of the  cross-correlations of the fluctuations
at different  sight lines across images produced at different
frequencies holds the possibility of allowing us to probe the two
point correlation function (or power spectrum) at high redshifts. We
have calculated the expected cross-correlations for two currently
acceptable CDM models and find it to be in the range $10^{-7} {\rm K^2}$
to $10^{-8} {\rm K^2}$ at $\nu \sim 320 {\rm MHz}$ for separations in
sight lines and frequencies such that the corresponding spatial
separation is in the range $10 \, {\rm Mpc}$ and $40 \, {\rm
Mpc}$. The cross-correlations are expected to be larger at smaller
scales where the linear theory cannot be applied.  
Our results show that both the isotropic background
(Figure~\ref{fig:2}) and its  fluctuations (Figure~\ref{fig:3} and
~\ref{fig:4}) can  be detected by GMRT which is the largest telescope
operating at meter waves at present, provided this signal can be
distinguished from other sources which contribute to the low frequency
background radiation. 

The biggest obstacles in detecting the HI contribution  are the
galactic and extra-galactic foregrounds,  both of which are many
orders larger than  the signal we want to detect.  The fact that both
those sources of contamination emit continuum radiation while 
the HI contribution is from  individual  damped Lyman-$\alpha$ clouds each
of which emits a spectral line  with a relatively small velocity width
keeps alive the possibility  of being able to distinguish this signal
from the contamination.  We have, in this paper, considered one possible
approach which might allow us to model and subtract the
foreground along any line of sight. More work is needed in this
direction  and work is currently underway in investigating  other
viable possibilities for foreground removal. 

\acknowledgments
All the authors would like to thank Jayaram Chengalur
for many useful discussions on issues related to the GMRT and
foregrounds.

\newpage 
\begin{figure}
\figurenum{1}
\includegraphics[angle=-90, width=0.5 \textwidth]{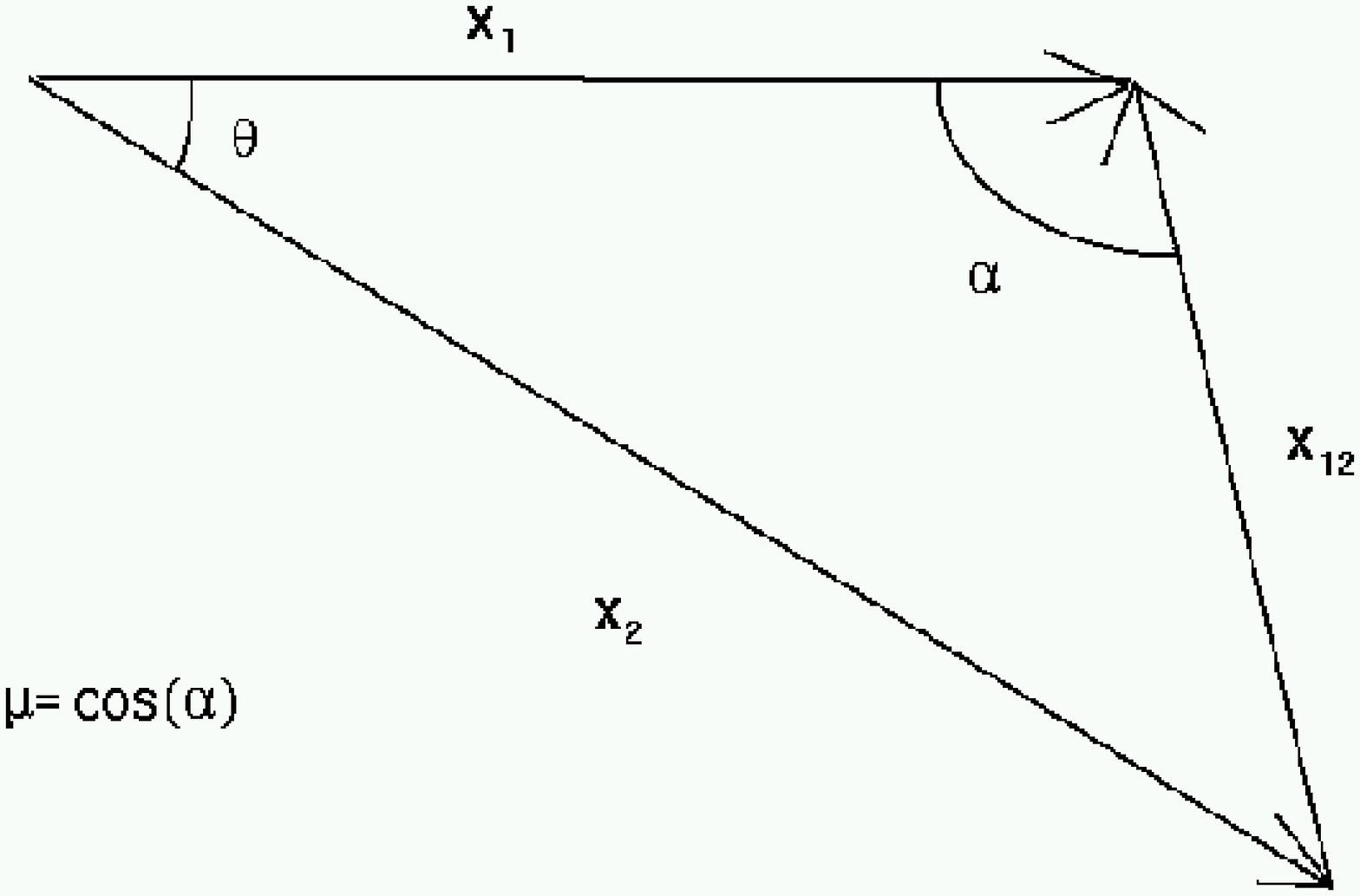} 
\caption{This shows the various distances and angles involved in
calculating $\w(\nu_1,\nu_2,\theta)$, where $\nu_1$ can be converted
to $x_1$ using equation (\ref{eq:a1}). \label{fig:1}}
\end{figure}

\begin{figure}
\figurenum{2}
\includegraphics[angle=-90, width=0.8 \textwidth]{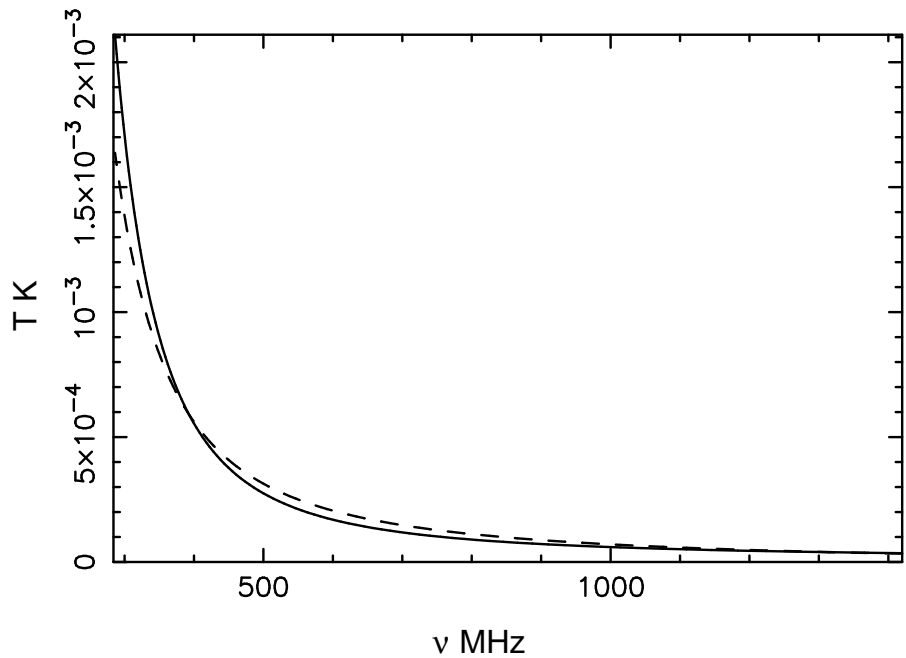}
\caption{This shows the isotropic part of the brightness temperature
at different frequencies. The solid line shows model (I) and the
dashed line model (II) of section 3.} 
\label{fig:2}
\end{figure}

\begin{figure}[b!]
\figurenum{3}
\includegraphics[angle=-90, width=0.45\textwidth]{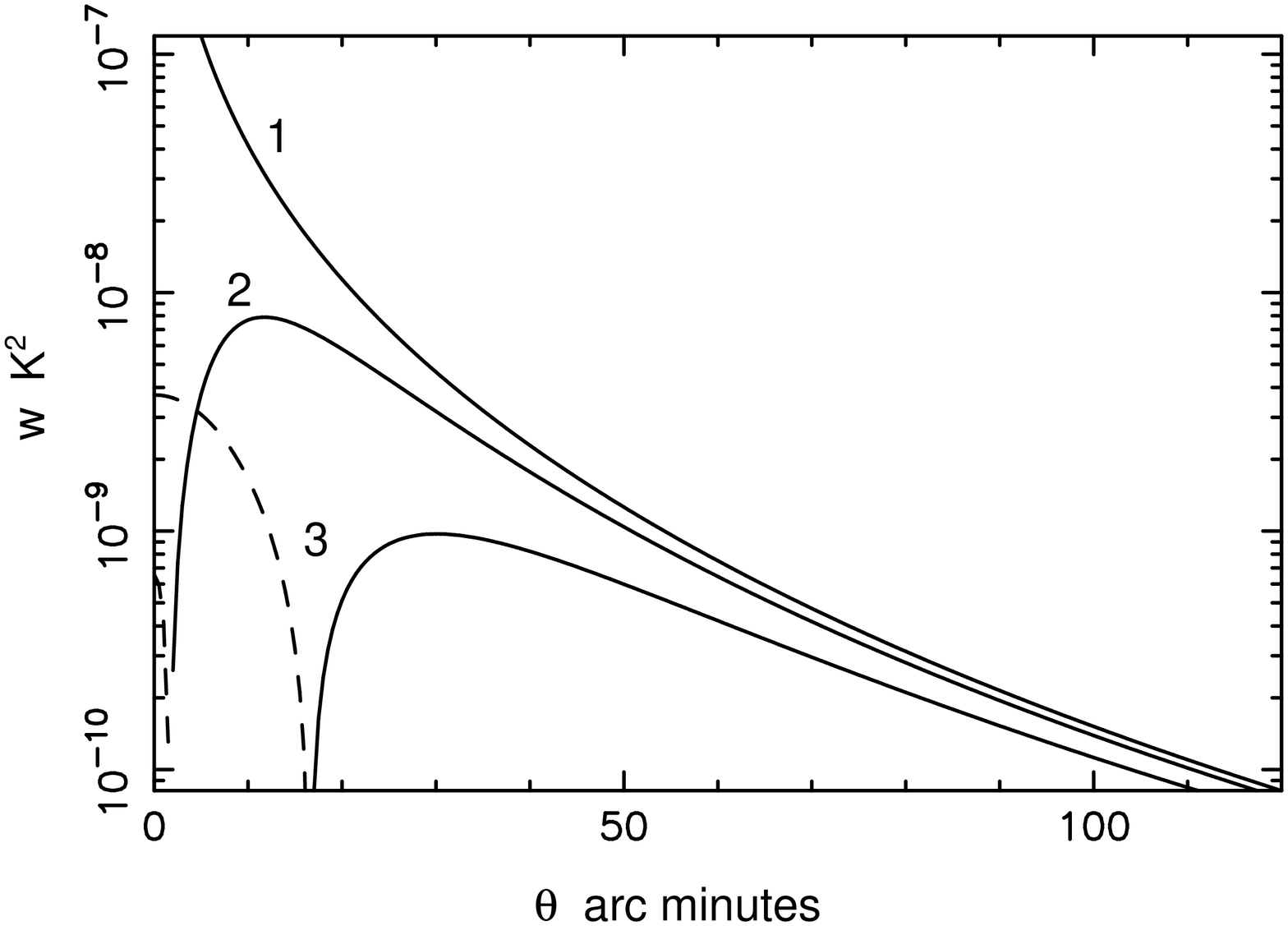}
\includegraphics[angle=-90, width=0.45\textwidth]{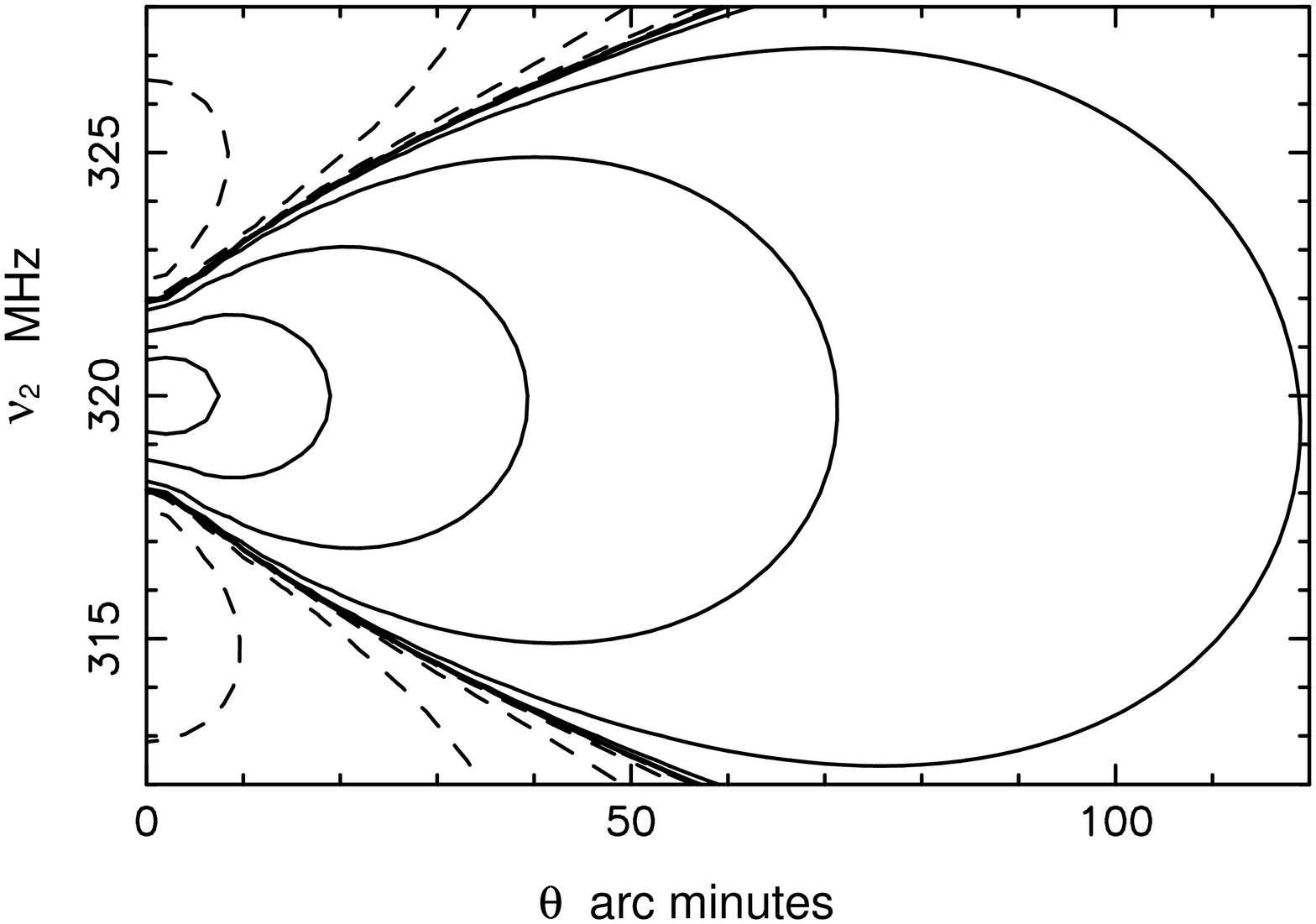}
\caption{For $\Omega_{m0}=1, h=0.5$, (a) shows
$\w(\nu_1,\nu_2,\theta)$ in ${\rm K}^2$  vs. $\theta$  with $\nu_1=320
{\rm MHz}$ and curves (1),(2) and (3) corresponding to   $\nu_2=320
{\rm MHz}$, $322 {\rm MHz}$  and $324 {\rm MHz}$ respectively;
 (b) shows contours of
equal $\w(\nu_1,\nu_2,\theta)$ at logarithmic intervals of $\w$ with  
$\nu_1=320 {\rm MHz}$. Here 
$1 {\rm MHz}=8.9 {\rm Mpc}$ and $1^{'}=1.8 {\rm Mpc}$.   The dashed
lines show negative values of $w$. } 
\label{fig:3}
\end{figure}

\begin{figure}[t!]
\figurenum{4}
\includegraphics[angle=-90, width=0.45\textwidth]{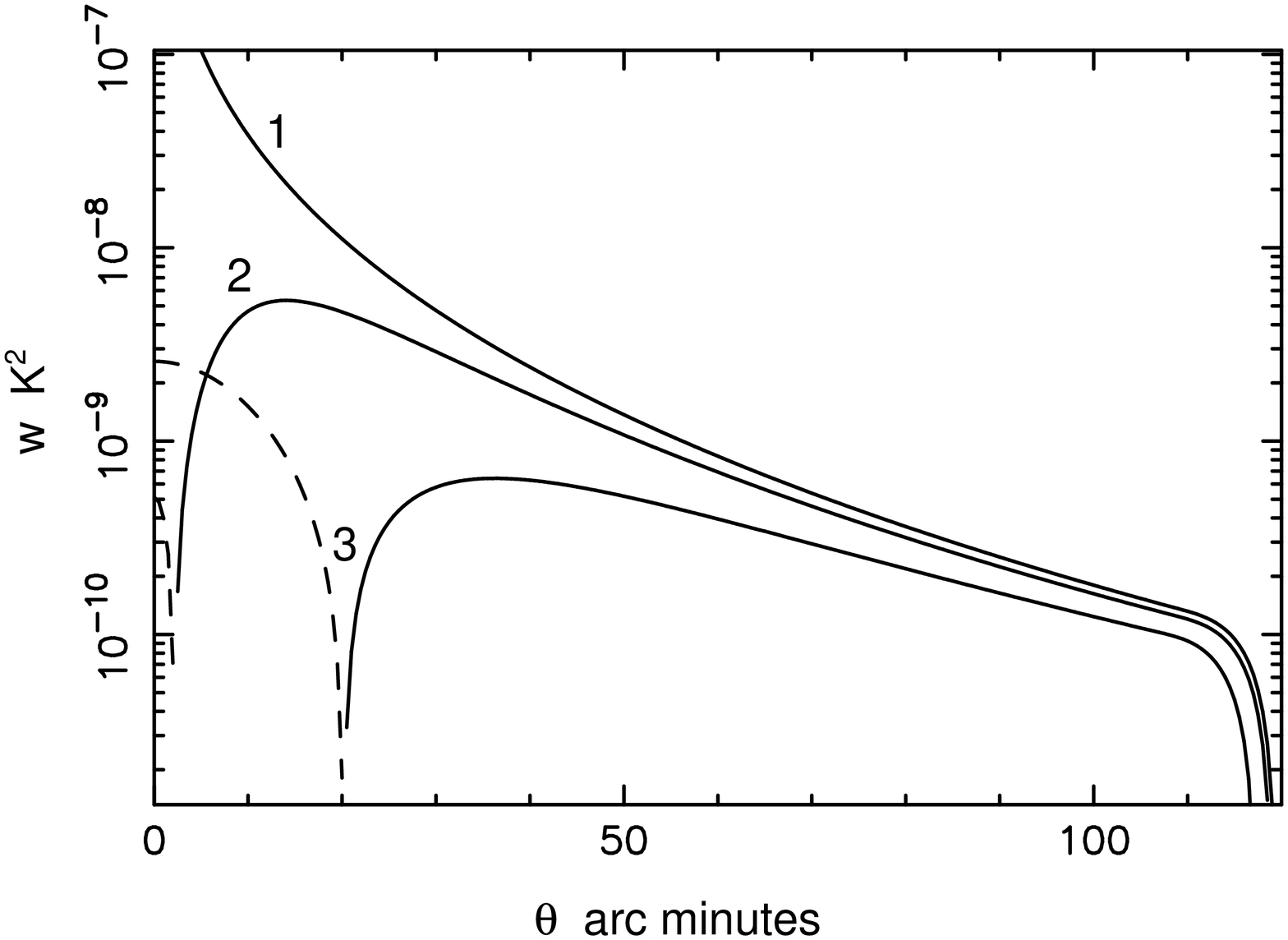}
\includegraphics[angle=-90, width=0.45\textwidth]{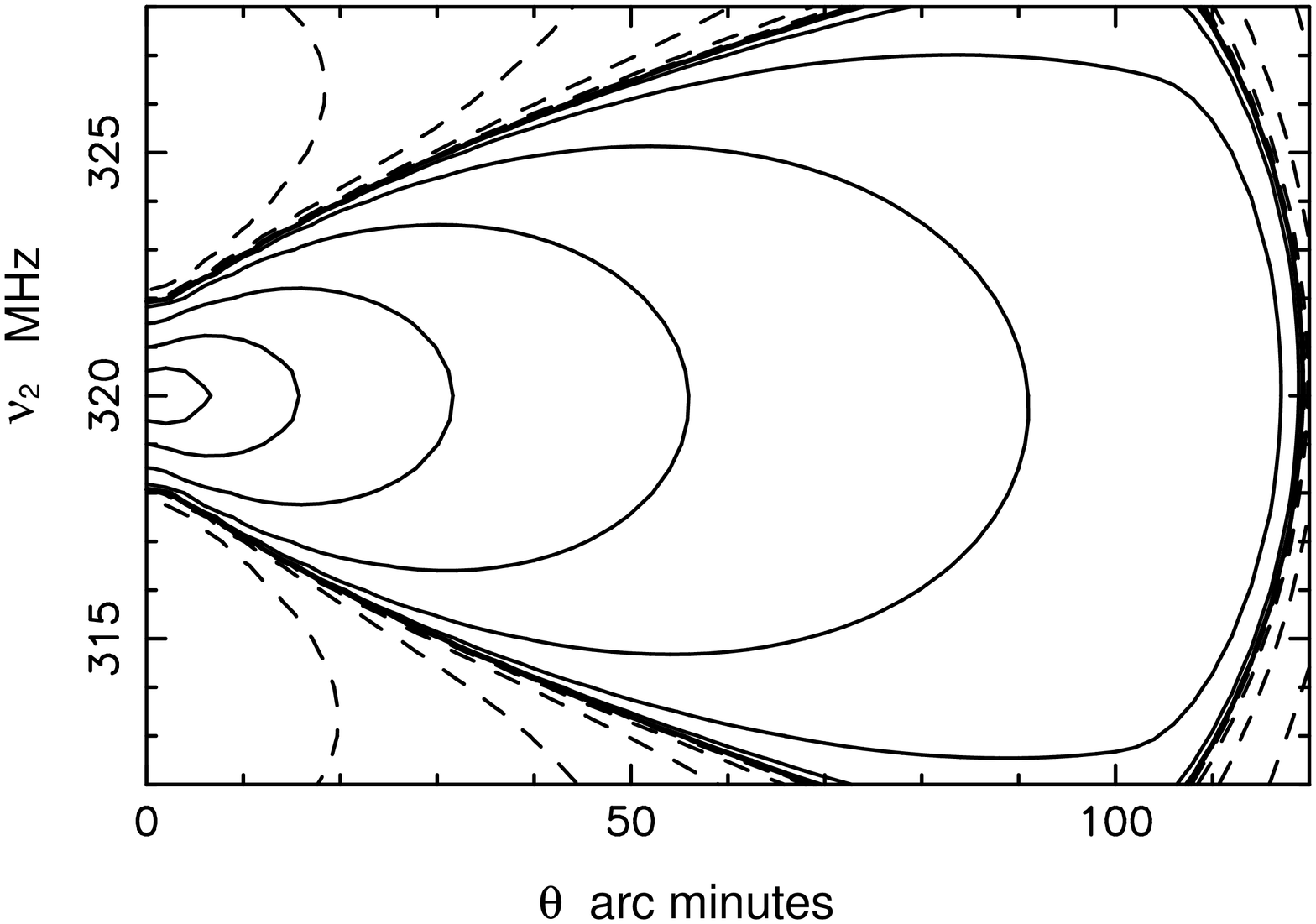}
\caption{For $\Omega_{m0}=0.3, \lambda_0=0.7,  h=0.5$, (a) shows
$\w(\nu_1,\nu_2,\theta)$ in ${\rm K}^2$  vs. $\theta$  with $\nu_1=320
{\rm MHz}$ and curves (1),(2) and (3) corresponding to   $\nu_2=320
{\rm MHz}$, $322 {\rm MHz}$  and $324 {\rm MHz}$ respectively;
 (b) shows contours of
equal $\w(\nu_1,\nu_2,\theta)$ at logarithmic intervals of $\w$ with  
$\nu_1=320 {\rm MHz}$. Here 
$1 {\rm MHz}=16.0 {\rm Mpc}$ and $1^{'}=2.7 {\rm Mpc}$.   The dashed
lines show negative values of $w$. } 
\label{fig:4}
\end{figure}

\end{document}